# Two-party secure semiquantum summation against the collective-dephasing noise

Tian-Yu Ye*, Tian-Jie Xu, Mao-Jie Geng, Ying Chen

College of Information & Electronic Engineering, Zhejiang Gongshang University, Hangzhou 310018, P.R.China

**Abstract:** In this paper, we propose a two-party semiquantum summation protocol, where two classical users can accomplish the summation of their private binary sequences with the assistance of a quantum semi-honest third party (TP). The term 'semi-honest' implies that TP cannot conspire with others but is able to implement all kinds oof attacks. This protocol employs logical qubits as traveling particles to overcome the negative influence of collective-dephasing noise and needn't make any two parties pre-share a random secret key. The security analysis turns out that this protocol can effectively prevent the outside attacks from Eve and the participant attacks from TP. Moreover, TP has no knowledge about the summation results.

## 1 Introduction

As a fundamental branch of quantum secure computation, the problem of secure quantum summation can be described as follows: $n$ users, $P_1, P_2, \ldots, P_n$, who possess the private inputs $X_1, X_2, \ldots, X_n$, respectively, want to securely compute the summation of $X_1, X_2, \ldots, X_n$, i.e., $sum(X_1, X_2, \ldots, X_n)$, on the basis that none of $X_1, X_2, \ldots, X_n$ is leaked out without being discovered. Recently, many researchers have shown their great interests on secure quantum summation so that it has been greatly developed. For instance, Heinrich not only introduced it into integration [1] but also considered quantum Boolean summation with repetitions under the worst-average setting [2]. Moreover, numerous secure quantum summation schemes [3-16] have appeared. However, each of these quantum summation schemes in Refs.[3-16] requires all parties to have complete quantum capabilities.

Recently, based on the well-known BB84 protocol [17], Boyer *et al.* [18-19] suggested the completely novel concept named as semi-quantum cryptography which doesn't have the demand that all parties must possess full quantum capabilities. In a semi-quantum cryptography protocol, the classical party is always restricted within the four operations, i.e., sending the qubits, reordering the qubits, generating the qubits in the Z basis (i.e., $\{|0\rangle, |1\rangle\}$) and measuring the qubits in the Z basis. [18-19] It is apparent that in a semi-quantum cryptography scheme, the classical party is free from the preparation and measurement of quantum superposition states and quantum entangled states. An important question naturally arises: whether there exists semiquantum summation or not? In the year of 2021, Zhang *et al.* [20] gave a positive answer to this question by putting forward the first semiquantum summation scheme. Note that Zhang *et al.*'s scheme in Ref.[20] is the only semiquantum summation scheme at present. But it doesn't consider the negative influence of noise so that it is only adaptive for the ideal noiseless quantum channel. In practice, photons are inevitably affected by the fluctuation of the birefringence in optical fiber, hence the negative influence of noise cannot be ignored. The channel noise can be considered as the collective noise, because photons travel inside a time window which is shorter than the variation of noise and are influenced by the same noise. [21] Hence, how to design a semiquantum summation scheme feasible for the collective noise quantum channel is urgent to solve.

Based on the above analysis, this paper concentrates on designing a semiquantum summation scheme immune to the collective-dephasing noise, which is one of the main kinds of collective noise. Compared with Zhang *et al.*'s semiquantum summation scheme [20], our scheme is more practical in reality, since it can resist the collective-dephasing noise; and our scheme has better privacy, since TP has no knowledge about the summation results.

## 2 Preliminary knowledge

$|0\rangle$ and $|1\rangle$ are the horizontal polarization and the vertical polarization of photon, respectively. When facing with the collective-dephasing noise, the former is kept unchanged, while the latter is turned into $e^{i\varphi}|1\rangle$, where $\varphi$ is the parameter of collective-dephasing noise fluctuating with time. [22] $|0_{dp}\rangle = |01\rangle$ and $|1_{dp}\rangle = |10\rangle$ are two logical qubits immune to the collective-dephasing noise. [22] Naturally, $|\pm_{dp}\rangle = \frac{1}{\sqrt{2}}(|0_{dp}\rangle \pm |1_{dp}\rangle) = \frac{1}{\sqrt{2}}(|01\rangle \pm |10\rangle)$ are also resistant against this kind of noise. [23] It is apparent that $Z_{dp} = \{|0_{dp}\rangle, |1_{dp}\rangle\}$ and $X_{dp} = \{|+_{dp}\rangle, |-_{dp}\rangle\}$ are two logical measuring bases under this kind of noise.

In addition, obviously, the four logical Bell entangled states defined as Eqs.(1-4) [24], are also resistant against this kind of noise, where $\left|\phi^{\pm}\right\rangle=\frac{1}{\sqrt{2}}(|00\rangle\pm|11\rangle)$ and $\left|\psi^{\pm}\right\rangle=\frac{1}{\sqrt{2}}(|01\rangle\pm|10\rangle)$ are four Bell entangled states. After being imposed with twice Bell state measurements on the first and the third physical qubits and on the second and the fourth physical qubits, respectively, these four logical Bell states can be clearly discriminated among each other. [24] In this paper, for the sake of convenience, this kind of quantum measurement is simply called as double Bell basis measurement.

$$\left|\Phi_{dp}^{+}\right\rangle_{1234} = \frac{1}{\sqrt{2}}\left(\left|+_{dp}\right\rangle\left|+_{dp}\right\rangle+\left|-_{dp}\right\rangle\left|-_{dp}\right\rangle\right)_{1234} = \frac{1}{\sqrt{2}}\left(\left|0_{dp}\right\rangle\left|0_{dp}\right\rangle+\left|1_{dp}\right\rangle\left|1_{dp}\right\rangle\right)_{1234} = \frac{1}{\sqrt{2}}\left(|01\rangle|01\rangle+|10\rangle|10\rangle\right)_{1234}$$
$$= \frac{1}{\sqrt{2}}\left(|00\rangle|11\rangle+|11\rangle|00\rangle\right)_{1324} = \frac{1}{\sqrt{2}}\left(\left|\phi^{+}\right\rangle\left|\phi^{+}\right\rangle-\left|\phi^{-}\right\rangle\left|\phi^{-}\right\rangle\right)_{1324}, \tag{1}$$

$$\left|\Phi_{dp}^{-}\right\rangle_{1234} = \frac{1}{\sqrt{2}}\left(\left|+_{dp}\right\rangle\left|-_{dp}\right\rangle+\left|-_{dp}\right\rangle\left|+_{dp}\right\rangle\right)_{1234} = \frac{1}{\sqrt{2}}\left(\left|0_{dp}\right\rangle\left|0_{dp}\right\rangle-\left|1_{dp}\right\rangle\left|1_{dp}\right\rangle\right)_{1234} = \frac{1}{\sqrt{2}}\left(|01\rangle|01\rangle-|10\rangle|10\rangle\right)_{1234}$$
$$= \frac{1}{\sqrt{2}}\left(|00\rangle|11\rangle-|11\rangle|00\rangle\right)_{1324} = \frac{1}{\sqrt{2}}\left(\left|\phi^{-}\right\rangle\left|\phi^{+}\right\rangle-\left|\phi^{+}\right\rangle\left|\phi^{-}\right\rangle\right)_{1324}, \tag{2}$$

$$\left|\Psi_{dp}^{+}\right\rangle_{1234} = \frac{1}{\sqrt{2}}\left(\left|+_{dp}\right\rangle\left|+_{dp}\right\rangle-\left|-_{dp}\right\rangle\left|-_{dp}\right\rangle\right)_{1234} = \frac{1}{\sqrt{2}}\left(\left|0_{dp}\right\rangle\left|1_{dp}\right\rangle+\left|1_{dp}\right\rangle\left|0_{dp}\right\rangle\right)_{1234} = \frac{1}{\sqrt{2}}\left(|01\rangle|10\rangle+|10\rangle|01\rangle\right)_{1234}$$
$$= \frac{1}{\sqrt{2}}\left(|01\rangle|10\rangle+|10\rangle|01\rangle\right)_{1324} = \frac{1}{\sqrt{2}}\left(\left|\psi^{+}\right\rangle\left|\psi^{+}\right\rangle-\left|\psi^{-}\right\rangle\left|\psi^{-}\right\rangle\right)_{1324}, \tag{3}$$

$$\left|\Psi_{dp}^{-}\right\rangle_{1234} = \frac{1}{\sqrt{2}}\left(\left|-_{dp}\right\rangle\left|+_{dp}\right\rangle-\left|+_{dp}\right\rangle\left|-_{dp}\right\rangle\right)_{1234} = \frac{1}{\sqrt{2}}\left(\left|0_{dp}\right\rangle\left|1_{dp}\right\rangle-\left|1_{dp}\right\rangle\left|0_{dp}\right\rangle\right)_{1234} = \frac{1}{\sqrt{2}}\left(|01\rangle|10\rangle-|10\rangle|01\rangle\right)_{1234}$$
$$= \frac{1}{\sqrt{2}}\left(|01\rangle|10\rangle-|10\rangle|01\rangle\right)_{1324} = \frac{1}{\sqrt{2}}\left(\left|\psi^{-}\right\rangle\left|\psi^{+}\right\rangle-\left|\psi^{+}\right\rangle\left|\psi^{-}\right\rangle\right)_{1324}. \tag{4}$$

According to Eq.(1) and Eq.(3), it has

$$\left|+_{dp}\right\rangle_{12}\left|+_{dp}\right\rangle_{34} = \frac{1}{\sqrt{2}}\left(\left|\Phi_{dp}^{+}\right\rangle_{1234}+\left|\Psi_{dp}^{+}\right\rangle_{1234}\right), \tag{5}$$

which means that if $\left|+_{dp}\right\rangle_{12}\left|+_{dp}\right\rangle_{34}$ is performed with double Bell basis measurement, it will be collapsed into $\left|\phi^{+}\right\rangle_{13}\left|\phi^{+}\right\rangle_{24}$, $\left|\phi^{-}\right\rangle_{13}\left|\phi^{-}\right\rangle_{24}$, $\left|\psi^{+}\right\rangle_{13}\left|\psi^{+}\right\rangle_{24}$ or $\left|\psi^{-}\right\rangle_{13}\left|\psi^{-}\right\rangle_{24}$ with equal probability. Moreover, it can be obtained from Eqs.(1-4) that

$$\left|0_{dp}\right\rangle_{12}\left|0_{dp}\right\rangle_{34} = \frac{1}{\sqrt{2}}\left(\left|\Phi_{dp}^{+}\right\rangle_{1234}+\left|\Phi_{dp}^{-}\right\rangle_{1234}\right), \tag{6}$$

$$\left|0_{dp}\right\rangle_{12}\left|1_{dp}\right\rangle_{34} = \frac{1}{\sqrt{2}}\left(\left|\Psi_{dp}^{+}\right\rangle_{1234}+\left|\Psi_{dp}^{-}\right\rangle_{1234}\right), \tag{7}$$

$$\left|1_{dp}\right\rangle_{12}\left|0_{dp}\right\rangle_{34} = \frac{1}{\sqrt{2}}\left(\left|\Psi_{dp}^{+}\right\rangle_{1234}-\left|\Psi_{dp}^{-}\right\rangle_{1234}\right), \tag{8}$$

$$\left|1_{dp}\right\rangle_{12}\left|1_{dp}\right\rangle_{34} = \frac{1}{\sqrt{2}}\left(\left|\Phi_{dp}^{+}\right\rangle_{1234}-\left|\Phi_{dp}^{-}\right\rangle_{1234}\right). \tag{9}$$

## 3 The proposed two-party semiquantum summation protocol

Suppose that there are two classical users with limited quantum capabilities, Alice and Bob. Alice's private binary string is denoted as

$$X = (x_1, x_2, \cdots, x_n), \tag{10}$$

while Bob's private binary string is represented as

$$Y = (y_1, y_2, \cdots, y_n). \tag{11}$$

Here, $x_j, y_j \in \{0,1\}$, $j = 1, 2, \cdots, n$. Alice and Bob want to calculate the modulo 2 summation of their private binary strings over the collective-dephasing noise quantum channel with the aid of a semi-honest third party (TP). A semi-honest TP is supposed to have the ability to perform all kinds of attacks but is not allowed to collude with anyone else [25]. A genuine two-user secure semiquantum summation protocol with a semi-honest TP should satisfy the following requirements [5]:

①Correctness. The summation result of two users' private binary strings should be correct;

②Security. Two users' private binary strings cannot be leaked out to an outside eavesdropper without being detected.

③Privacy. Each user's private binary string should be kept secret from TP.

Recently, Zhang *et al.* [26] designed a semiquantum key distribution (SQKD) protocol robust against the collective-dephasing noise; Lin *et al.* [27] put forward a semiquantum private comparison (SQPC) protocol by using the quantum states $|+\rangle$, where $|+\rangle = \frac{1}{\sqrt{2}}(|0\rangle + |1\rangle)$. Inspired by Refs.[26,27], we design the following semiquantum summation protocol to compute the modulo 2 summation of Alice and Bob's private binary strings over the collective-dephasing noise quantum channel.

**Step 1:** TP generates $2n(4+r+d+\delta) = 2nq$ particles all in the state of $|+_{dp}\rangle$. Here, $r, d$ are integers greater than 0; $\delta$ is some fixed parameter greater than 0; and $q = (4+r+d+\delta)$. Then, TP divides these particles into two sequences, $S_1 = \{s_1^1, s_1^2, \cdots, s_1^{nq}\}$ and $S_2 = \{s_2^1, s_2^2, \cdots, s_2^{nq}\}$, where $s_1^i$ and $s_2^i$ denote the $i^{th}$ particle in $S_1$ and $S_2$, respectively, and $i = 1, 2, \cdots, nq$. Finally, TP sends the particles of $S_1$ and $S_2$ to Alice and Bob one by one, respectively. Except the first particle, TP sends out the next one only after obtaining the previous one.

**Step 2:** For each received particle in $S_1$ ($S_2$), Alice (Bob) immediately randomly performs one of the following two operations: directly reflecting it back to TP with no disturbance (i.e., the CTRL operation) or measuring it with the $Z_{dp}$ basis and resending the same state as found to TP (i.e., the SIFT operation). After Alice's (Bob's) operations, $S_1$ ($S_2$) is turned into $S_1'$ ($S_2'$). TP stores $S_1'$ ($S_2'$) in a quantum memory.

**Step 3:** TP picks out the $i^{th}$ particle in $S_1'$ and the $i^{th}$ particle in $S_2'$ to form the $i^{th}$ particle group, where $i = 1, 2, \cdots, nq$. For checking the transmission security towards an outside eavesdropper Eve, TP randomly chooses $nr$ groups from these particle groups, and tells Alice and Bob the positions of the chosen particle groups. Among these chosen particle groups, Alice (Bob) tells TP the positions of particles where she (he) chose the CTRL operations, the positions of particles where she (he) chose the SIFT operations as well as her (his) measurement results.

For the particles on which Alice (Bob) performed the CTRL operations, TP measures them with the $X_{dp}$ basis. TP computes the error rate of CTRL particles by judging whether her measurement results are $|+_{dp}\rangle$ or not. The protocol is kept on only when the transmission of CTRL particles is secure.

For the particles on which Alice (Bob) performed the SIFT operations, TP measures them with the $Z_{dp}$ basis. TP computes the error rate of SIFT particles by judging whether her measurement results are identical to Alice's (Bob's) corresponding measurement results or not. The protocol is kept on only when the transmission of SIFT particles is secure.

**Step 4:** Alice and Bob ask TP to perform the double Bell basis measurement on each of the remaining $n(4+d+\delta)$ particle groups and announce them her corresponding measurement result. After confirming that TP have announced them all of the double Bell basis measurements on the remaining $n(4+d+\delta)$ particle groups, Alice and Bob randomly choose $nd$ groups from the remaining $n(4+d+\delta)$ particle groups to check the honesty of TP. For the particle group on which both Alice and Bob performed the CTRL operations, if TP's measurement result is not $|\phi^+\rangle_{13}|\phi^+\rangle_{24}, |\phi^-\rangle_{13}|\phi^-\rangle_{24}, |\psi^+\rangle_{13}|\psi^+\rangle_{24}$ or $|\psi^-\rangle_{13}|\psi^-\rangle_{24}$, according to Eq.(5), Alice and Bob will think that TP is dishonest. For the particle group on which both Alice and Bob performed the SIFT operations, Alice and Bob check whether TP's measurement result satisfies Eqs.(6-9) or not; if the result is negative, Alice and Bob will think that TP is dishonest. For example, if Alice and Bob's measurement results are $|0_{dp}\rangle_{12}$ and $|0_{dp}\rangle_{34}$, respectively, according to Eq.(6), TP's measurement result should be $|\phi^+\rangle_{13}|\phi^+\rangle_{24}, |\phi^-\rangle_{13}|\phi^-\rangle_{24}, |\phi^+\rangle_{13}|\phi^-\rangle_{24}$ or $|\phi^-\rangle_{13}|\phi^+\rangle_{24}$; otherwise, Alice and Bob will conclude that TP is dishonest. The protocol is kept on only when TP is found to be honest in the end.

**Step 5:** For the remaining $n(4+\delta)$ particle groups, Alice (Bob) tells Bob (Alice) the positions of particles where she (he)

performed the SIFT operations. Note that there are $n(4+\delta) \times \frac{1}{4} = n + \frac{n\delta}{4}$ particle groups where both Alice and Bob performed the SIFT operations. Alice and Bob choose the first $n$ ones from these $n + \frac{n\delta}{4}$ particle groups to generate their private keys. The private key generating rule is that: if Alice's (Bob's) measurement result on the corresponding particle from the $j^{th}$ particle group is $|0_{dp}\rangle$, her (his) $j^{th}$ bit of private key, $k_j^a$ ($k_j^b$), will be 0; and if Alice's (Bob's) measurement result on it is $|1_{dp}\rangle$, $k_j^a$ ($k_j^b$) will be 1. Here, $j = 1, 2, \cdots, n$; $K_A = [k_1^a, k_2^a, \cdots, k_n^a]$ and $K_B = [k_1^b, k_2^b, \cdots, k_n^b]$ are Alice and Bob's private keys, respectively. Then, Alice (Bob) also derives a private bit string $C_T$ from TP's measurement results on these $n$ particle groups according to the following rule: if TP's measurement result on the $j^{th}$ particle group is $|\phi^+\rangle_{13}|\phi^+\rangle_{24}, |\phi^-\rangle_{13}|\phi^-\rangle_{24}, |\phi^+\rangle_{13}|\phi^-\rangle_{24}$ or $|\phi^-\rangle_{13}|\phi^+\rangle_{24}$, $k_j^t$ will be 0; and if it is $|\psi^+\rangle_{13}|\psi^+\rangle_{24}, |\psi^-\rangle_{13}|\psi^-\rangle_{24}, |\psi^+\rangle_{13}|\psi^-\rangle_{24}$ or $|\psi^-\rangle_{13}|\psi^+\rangle_{24}$, $k_j^t$ will be 1. Here, $k_j^t$ is the $j^{th}$ bit of $C_T$, and $j = 1, 2, \cdots, n$. Afterward, Alice (Bob) calculates $c_j^a = k_j^a \oplus x_j$ ($c_j^b = k_j^b \oplus y_j$), where $\oplus$ is the modulo 2 summation, and $j = 1, 2, \cdots, n$. Alice (Bob) sends $C_A$ ($C_B$) to Bob (Alice) via the classical channel, where $C_A = [c_1^a, c_2^a, \cdots, c_n^a]$ ($C_B = [c_1^b, c_2^b, \cdots, c_n^b]$). Finally, Alice (Bob) calculates $r_j = c_j^a \oplus c_j^b \oplus k_j^t$, where $j = 1, 2, \cdots, n$, and obtains the summation result $R$, where $R = [r_1, r_2, \cdots, r_n]$.

For the sake of clarity, the relations among different parameters for summation are listed in Table 1.

Table 1　Relations among different parameters for summation

| $x_j$ | $y_j$ | Alice's measurement result on the corresponding particle from the $j^{th}$ particle group | Bob's measurement result on the corresponding particle from the $j^{th}$ particle group | $k_j^a$ | $k_j^b$ | $c_j^a$ | $c_j^b$ | TP's double Bell basis measurement result on the $j^{th}$ particle group | $k_j^t$ | $r_j$ |
|---|---|---|---|---|---|---|---|---|---|---|
| 0 | 0 | $|0_{dp}\rangle$ | $|0_{dp}\rangle$ | 0 | 0 | 0 | 0 | $|\phi^+\rangle_{13}|\phi^+\rangle_{24}, |\phi^-\rangle_{13}|\phi^-\rangle_{24}, |\phi^+\rangle_{13}|\phi^-\rangle_{24}, |\phi^-\rangle_{13}|\phi^+\rangle_{24}$ | 0 | 0 |
| | | $|0_{dp}\rangle$ | $|1_{dp}\rangle$ | 0 | 1 | 0 | 1 | $|\psi^+\rangle_{13}|\psi^+\rangle_{24}, |\psi^-\rangle_{13}|\psi^-\rangle_{24}, |\psi^+\rangle_{13}|\psi^-\rangle_{24}, |\psi^-\rangle_{13}|\psi^+\rangle_{24}$ | 1 | 0 |
| | | $|1_{dp}\rangle$ | $|0_{dp}\rangle$ | 1 | 0 | 1 | 0 | $|\psi^+\rangle_{13}|\psi^+\rangle_{24}, |\psi^-\rangle_{13}|\psi^-\rangle_{24}, |\psi^+\rangle_{13}|\psi^-\rangle_{24}, |\psi^-\rangle_{13}|\psi^+\rangle_{24}$ | 1 | 0 |
| | | $|1_{dp}\rangle$ | $|1_{dp}\rangle$ | 1 | 1 | 1 | 1 | $|\phi^+\rangle_{13}|\phi^+\rangle_{24}, |\phi^-\rangle_{13}|\phi^-\rangle_{24}, |\phi^+\rangle_{13}|\phi^-\rangle_{24}, |\phi^-\rangle_{13}|\phi^+\rangle_{24}$ | 0 | 0 |
| 0 | 1 | $|0_{dp}\rangle$ | $|0_{dp}\rangle$ | 0 | 0 | 0 | 1 | $|\phi^+\rangle_{13}|\phi^+\rangle_{24}, |\phi^-\rangle_{13}|\phi^-\rangle_{24}, |\phi^+\rangle_{13}|\phi^-\rangle_{24}, |\phi^-\rangle_{13}|\phi^+\rangle_{24}$ | 0 | 1 |
| | | $|0_{dp}\rangle$ | $|1_{dp}\rangle$ | 0 | 1 | 0 | 0 | $|\psi^+\rangle_{13}|\psi^+\rangle_{24}, |\psi^-\rangle_{13}|\psi^-\rangle_{24}, |\psi^+\rangle_{13}|\psi^-\rangle_{24}, |\psi^-\rangle_{13}|\psi^+\rangle_{24}$ | 1 | 1 |
| | | $|1_{dp}\rangle$ | $|0_{dp}\rangle$ | 1 | 0 | 1 | 1 | $|\psi^+\rangle_{13}|\psi^+\rangle_{24}, |\psi^-\rangle_{13}|\psi^-\rangle_{24}, |\psi^+\rangle_{13}|\psi^-\rangle_{24}, |\psi^-\rangle_{13}|\psi^+\rangle_{24}$ | 1 | 1 |
| | | $|1_{dp}\rangle$ | $|1_{dp}\rangle$ | 1 | 1 | 1 | 0 | $|\phi^+\rangle_{13}|\phi^+\rangle_{24}, |\phi^-\rangle_{13}|\phi^-\rangle_{24}, |\phi^+\rangle_{13}|\phi^-\rangle_{24}, |\phi^-\rangle_{13}|\phi^+\rangle_{24}$ | 0 | 1 |
| 1 | 0 | $|0_{dp}\rangle$ | $|0_{dp}\rangle$ | 0 | 0 | 1 | 0 | $|\phi^+\rangle_{13}|\phi^+\rangle_{24}, |\phi^-\rangle_{13}|\phi^-\rangle_{24}, |\phi^+\rangle_{13}|\phi^-\rangle_{24}, |\phi^-\rangle_{13}|\phi^+\rangle_{24}$ | 0 | 1 |
| | | $|0_{dp}\rangle$ | $|1_{dp}\rangle$ | 0 | 1 | 1 | 1 | $|\psi^+\rangle_{13}|\psi^+\rangle_{24}, |\psi^-\rangle_{13}|\psi^-\rangle_{24}, |\psi^+\rangle_{13}|\psi^-\rangle_{24}, |\psi^-\rangle_{13}|\psi^+\rangle_{24}$ | 1 | 1 |
| | | $|1_{dp}\rangle$ | $|0_{dp}\rangle$ | 1 | 0 | 0 | 0 | $|\psi^+\rangle_{13}|\psi^+\rangle_{24}, |\psi^-\rangle_{13}|\psi^-\rangle_{24}, |\psi^+\rangle_{13}|\psi^-\rangle_{24}, |\psi^-\rangle_{13}|\psi^+\rangle_{24}$ | 1 | 1 |
| | | $|1_{dp}\rangle$ | $|1_{dp}\rangle$ | 1 | 1 | 0 | 1 | $|\phi^+\rangle_{13}|\phi^+\rangle_{24}, |\phi^-\rangle_{13}|\phi^-\rangle_{24}, |\phi^+\rangle_{13}|\phi^-\rangle_{24}, |\phi^-\rangle_{13}|\phi^+\rangle_{24}$ | 0 | 1 |
| 1 | 1 | $|0_{dp}\rangle$ | $|0_{dp}\rangle$ | 0 | 0 | 1 | 1 | $|\phi^+\rangle_{13}|\phi^+\rangle_{24}, |\phi^-\rangle_{13}|\phi^-\rangle_{24}, |\phi^+\rangle_{13}|\phi^-\rangle_{24}, |\phi^-\rangle_{13}|\phi^+\rangle_{24}$ | 0 | 0 |
| | | $|0_{dp}\rangle$ | $|1_{dp}\rangle$ | 0 | 1 | 1 | 0 | $|\psi^+\rangle_{13}|\psi^+\rangle_{24}, |\psi^-\rangle_{13}|\psi^-\rangle_{24}, |\psi^+\rangle_{13}|\psi^-\rangle_{24}, |\psi^-\rangle_{13}|\psi^+\rangle_{24}$ | 1 | 0 |
| | | $|1_{dp}\rangle$ | $|0_{dp}\rangle$ | 1 | 0 | 0 | 1 | $|\psi^+\rangle_{13}|\psi^+\rangle_{24}, |\psi^-\rangle_{13}|\psi^-\rangle_{24}, |\psi^+\rangle_{13}|\psi^-\rangle_{24}, |\psi^-\rangle_{13}|\psi^+\rangle_{24}$ | 1 | 0 |

| | | | | | | | | |
|---|---|---|---|---|---|---|---|---|
| $\|1_{dp}\rangle$ | $\|1_{dp}\rangle$ | 1 | 1 | 0 | 0 | $\|\phi^+\rangle_{13}\|\phi^+\rangle_{24}, \|\phi^-\rangle_{13}\|\phi^-\rangle_{24}, \|\phi^+\rangle_{13}\|\phi^-\rangle_{24}, \|\phi^-\rangle_{13}\|\phi^+\rangle_{24}$ | 0 | 0 |

Obviously, TP needs to have full quantum capabilities. Both Alice and Bob only implement the following actions: ① measuring the particles with the $Z_{dp}$ basis; ② preparing the particles in the $Z_{dp}$ basis; ③ sending the particles. According to Refs.[18-19], the $Z$ basis is regarded to be classical, so the $Z_{dp}$ basis is naturally classical. Consequently, both Alice and Bob only need limited quantum capabilities. In other words, this protocol is a semiquantum summation protocol.

## 4 Correctness analysis

In the proposed protocol, after receiving $C_A$ ($C_B$) from Alice (Bob), Bob (Alice) calculates $r_j = c_j^a \oplus c_j^b \oplus k_j^t$, where $j = 1, 2, \cdots, n$. According to Eqs.(6-9), it is apparent that

$$k_j^a \oplus k_j^b \oplus k_j^t = 0. \tag{12}$$

Hence, it has

$$r_j = c_j^a \oplus c_j^b \oplus k_j^t = (k_j^a \oplus x_j) \oplus (k_j^b \oplus y_j) \oplus k_j^t = (x_j \oplus y_j) \oplus (k_j^a \oplus k_j^b \oplus k_j^t) = x_j \oplus y_j. \tag{13}$$

It can be concluded that the output correctness of the proposed protocol can be guaranteed.

## 5 Security analysis

**(1) Outside attack**

An outside attacker, Eve, may try her best to extract $X$ ($Y$) from $C_A$ ($C_B$). Apparently, she should obtain $K_A$ ($K_B$) beforehand. Without loss of generality, here take Eve's trying to get $K_A$ for example to analyze the outside attack.

**Measure-resend attack.** Eve measures the particles in $S_1$ from TP to Alice in Step 1 with the $Z_{dp}$ basis and sends the new particles in the same states she found to Alice. However, she will be caught as she has no access to Alice's choices of operations in Step 2. Concretely speaking, for one particular particle chosen for detection, the probability that Alice chooses the CTRL operation is $\frac{1}{2}$; hence, the probability that Eve can be caught is $\frac{1}{2} \times \frac{1}{2} = \frac{1}{4}$, since TP has a $\frac{1}{2}$ probability to get the wrong measurement result $\left|-_{dp}\right\rangle$ on the particle reflected by Alice. For $nr$ particle groups chosen for the security check in Step 3, the probability of Eve's being discovered becomes $1 - \left(\frac{3}{4}\right)^{nr}$, which will converge to 1 if the value of $nr$ is large enough.

**Intercept-resend attack.** Eve intercepts the particles in $S_1$ from TP to Alice in Step 1, sends the fake particle sequence $S_1^E$ she prepared in the $Z_{dp}$ basis beforehand to Alice; Eve intercepts $S_1^E$ after Alice's operations and sends $S_1$ to TP. For one particular particle chosen for detection, the probability that Alice chooses the SIFT operation is $\frac{1}{2}$; hence, the probability that Eve can be caught is $\frac{1}{2} \times \frac{1}{2} = \frac{1}{4}$, as there is a $\frac{1}{2}$ probability that Alice's measurement result on Eve's fake particle is not same to TP's measurement result on the genuine one. For $nr$ particle groups chosen for the security check in Step 3, the probability of Eve's being discovered becomes $1 - \left(\frac{3}{4}\right)^{nr}$.

**Double CNOT attack.** Eve may perform the first CNOT operations, defined as

$$CNOT = |00\rangle\langle00| + |01\rangle\langle01| + |11\rangle\langle10| + |10\rangle\langle11|, \tag{14}$$

on the particles in $S_1$ and her own ancillary particles in the state of $\left|0_{dp}\right\rangle$ in Step 1. Here, the first physical qubit of each particle in $S_1$ acts as the control qubit while two physical qubits of each ancillary particle play the role of target qubits. In order to escape the security check in Step 3, Eve has to perform the second CNOT operations on the particles in $S_1^{'}$ and her own ancillary particles in Step 2. Here, the first physical qubit of each particle in $S_1^{'}$ acts as the control qubit while two physical qubits of each ancillary particle play the role of target qubits. If Eve can know Alice's choices of operations in Step 2 exactly through the double CNOT attacks, she will further perform the SIFT operations on the particles where Alice has implemented the same operations, in order to obtain $K_A$ without being discovered. However, Eve cannot discriminate Alice's choices of operations at all. Concretely speaking, after the first CNOT operation, the composite system composed by one particle in $S_1$ and the ancillary particle $|0_{dp}\rangle$ is turned into

$$CNOT|+_{dp}\rangle_A |0_{dp}\rangle_E = \frac{1}{\sqrt{2}}(|0_{dp}\rangle_A |0_{dp}\rangle_E + |1_{dp}\rangle_A |1_{dp}\rangle_E) = |\Phi_{dp}^+\rangle_{AE}, \quad (15)$$

where the subscripts $A$ and $E$ represent Alice's particle and Eve's ancillary particle, respectively. If Alice chooses the CTRL operation, after Eve's second CNOT operation, the composite system will be changed into

$$CNOT|\Phi_{dp}^+\rangle_{AE} = |+_{dp}\rangle_A |0_{dp}\rangle_E. \quad (16)$$

If Alice chooses the SIFT operation and obtains the measurement result $|0_{dp}\rangle$ ($|1_{dp}\rangle$), according to Eq.(15), Eve's ancillary particle will be collapsed into $|0_{dp}\rangle$ ($|1_{dp}\rangle$). After Eve's second CNOT operation, the composite system is evolved into

$$CNOT|0_{dp}\rangle_A |0_{dp}\rangle_E = |0_{dp}\rangle_A |0_{dp}\rangle_E, \text{ if Alice's measurement result is } |0_{dp}\rangle, \quad (17)$$

$$CNOT|1_{dp}\rangle_A |1_{dp}\rangle_E = |1_{dp}\rangle_A |0_{dp}\rangle_E, \text{ if Alice's measurement result is } |1_{dp}\rangle. \quad (18)$$

According to Eqs.(16-18), no matter what operation Alice chooses in Step 2, after the second CNOT operation, Eve's ancillary particle is always in the state of $|0_{dp}\rangle$. In other words, Eve cannot discriminate Alice's choice of operation through her ancillary particle.

It is worthy of emphasizing that if Eve doesn't perform the second CNOT operation, Eve's CNOT attack will inevitably be detected in Step 3, as TP has a $\frac{1}{2}$ probability to get the wrong measurement result $|-_{dp}\rangle$ for the particle chosen for detection on which Alice have performed the CTRL operation, according to Eq.(1) and Eq.(15).

**Trojan horse attacks.** As the travelling particles are transmitted in a circular way, the Trojan horse attacks from Eve should be paid special attention to, such as the invisible photon eavesdropping attack [28] and the delay-photon Trojan horse attack [29-30]. It has been verified that the signal receiver can use a filter to resist the former attack and a photon number splitter (PNS) to defeat the latter attack [30-31].

**(2) Participant attack**

Gao et al. [32] firstly warned against the participant attack in the year of 2007 when designing a quantum cryptography protocol. It is natural that a $m$-user quantum summation protocol cannot resist the collusion attack from $m-1$ users. The reason lies in that: when $m-1$ users conspire together, they can easily derive the private input of the left user from the summation result. With regard to the proposed protocol, Alice (Bob) can easily deduce out $Y$ ($X$) from $R$. As a result, we only need to consider the participant attack from the semi-honest TP.

Apparently, if TP wants to derive $X$ ($Y$) from $C_A$ ($C_B$), she will need to get $K_A$ ($K_B$) beforehand. In order to achieve this goal, TP may launch the following attacks.

**Attack I:** TP may generate all particles in the $Z_{dp}$ basis and transmits them to Alice and Bob in Step 1; moreover, TP always announces the genuine double Bell basis measurement results to Alice and Bob in Step 4. In this way, if TP's cheating behavior successfully passes the honesty check in Step 4, she will easily obtain $K_A$ and $K_B$ in Step 5. However, TP's cheating behavior is inevitably discovered by Alice and Bob during the honesty check in Step 4. Concretely speaking, for one particle group chosen for TP's honesty check, according to Eqs.(5-9), if both Alice and Bob choose the CTRL operations, TP will be detected with the probability of $\frac{1}{2}$; if both Alice and Bob choose the SIFT operations, TP will be detected with the probability of 0; and if only one chooses the CTRL operation, TP will be automatically not detected, since there is no check for this case. Thus, for one particle group chosen for TP's honesty check, the probability that TP will be detected is $\frac{1}{2} \times \frac{1}{2} \times \frac{1}{2} = \frac{1}{8}$. For $nd$ particle groups chosen for TP's honesty check, the probability that TP will be detected is $1 - \left(\frac{7}{8}\right)^{nd}$, which will converge to 1 if the value of $nd$ is large enough.

**Attack II:** TP measures all particles in the remaining $n(4+d+\delta)$ particle groups with the $Z_{dp}$ basis instead of the double Bell basis, and announces the fake double Bell basis measurement results to Alice and Bob in Step 4, in hope of escaping the honesty check. However, TP's cheating behavior is undoubtedly detected as she cannot exactly know Alice and Bob's choices of operations in Step 2. Concretely speaking, for one particle group chosen for TP's honesty check, according to Eqs.(6-9), if TP's measurement result is $|0_{dp}\rangle|0_{dp}\rangle$ or $|1_{dp}\rangle|1_{dp}\rangle$, TP will randomly announce the fake measurement result $|\phi^+\rangle_{13}|\phi^+\rangle_{24}$, $|\phi^-\rangle_{13}|\phi^-\rangle_{24}$, $|\phi^+\rangle_{13}|\phi^-\rangle_{24}$ or $|\phi^-\rangle_{13}|\phi^+\rangle_{24}$; and if TP's measurement result is $|0_{dp}\rangle|1_{dp}\rangle$ or $|1_{dp}\rangle|0_{dp}\rangle$, TP will randomly announce the fake measurement result $|\psi^+\rangle_{13}|\psi^+\rangle_{24}$, $|\psi^-\rangle_{13}|\psi^-\rangle_{24}$, $|\psi^+\rangle_{13}|\psi^-\rangle_{24}$ or $|\psi^-\rangle_{13}|\psi^+\rangle_{24}$. As a result, if both Alice and Bob choose the SIFT operations, TP will be detected with the probability of 0; and if both Alice and Bob choose the CTRL operations, TP will be detected with the probability of $\frac{1}{2}$. Thus, for one particle group chosen for TP's honesty check, the probability that TP will be detected is $\frac{1}{2} \times \frac{1}{2} \times \frac{1}{2} = \frac{1}{8}$. For $nd$ particle groups chosen for TP's honesty check, the probability that TP will be detected is $1-\left(\frac{7}{8}\right)^{nd}$.

It can be concluded that TP's attack is inevitably discovered by the honesty check in Step 4 when she ties to extract Alice and Bob's private keys.

## 6 Discussions and conclusions

We firstly discuss the qubit efficiency [33] defined as

$$\eta = \frac{c}{p+v}, \tag{19}$$

where $c$ is the total number of classical bits for summation; $q$ is the total number of consumed qubits; and $v$ is the total number of classical bits needed for summation. Here, the classical bits consumed for eavesdropping check are not taken into account. In this protocol, both the length of Alice's private binary string and the length of Bob's private binary string are $n$, hence, we have $c = n$; TP needs to generate $2n(4+r+d+\delta) = 2nq$ particles all in the state of $|+_{dp}\rangle$, while both Alice and Bob need to prepare $\frac{nq}{2}$ particles in the $Z_{dp}$ basis when implementing the SIFT operations, hence, we have $p = 2nq \times 2 + \frac{nq}{2} \times 2 \times 2 = 6nq$; TP needs to publish Alice and Bob her double Bell basis measurement results on the $n$ particle groups used for summation, while Alice and Bob need to send $C_A$ and $C_B$ to the other via the classical channel, respectively, hence, we have $v = 6n$. Consequently, it can be obtained that $\eta = \frac{n}{6nq+6n} = \frac{1}{6(4+r+d+\delta)+6}$.

We compare this protocol with the only existing semiquantum summation protocol [20] in detail and list the comparison result in Table 2. In Table 2, with respect to this protocol, the two-qubit entangled state quantum resource refers to the state $|+_{dp}\rangle$; both TP's single-qubit measurements and communicants' single-qubit measurements refer to the $Z_{dp}$ basis measurements, as the $Z_{dp}$ basis measurement is composed by two composite $Z$ basis measurements; and TP's two-qubit entangled state measurements refer to the $X_{dp}$ basis measurements and the double Bell basis measurements. As for the protocol of Ref.[20], the single-qubit state quantum resource refers to the state $|+\rangle$; communicants' single-qubit measurements refer to the $Z$ basis measurements; TP's single-qubit measurements refer to the $X$ basis ($\{|+\rangle,|-\rangle\}$) measurements and the $Z$ basis measurements; and TP's three-qubit entangled state measurements refer to the GHZ-type basis measurements. In addition, in the protocol of Ref.[20], the length of the private binary string from $P_j$ is $n$, where $j = 1,2,3$; TP needs to generate $3n(32+r+d+\delta) = 3nt$ particles all in the state of $|+\rangle$, while $P_j$ needs to generate $\frac{nt}{2}$ particles in the $Z$ basis when choosing the SIFT operations; and $P_j$ needs to send $C_j$ to TP through the classical channel; hence, we have $c = n$, $p = 3nt + \frac{nt}{2} \times 3 = \frac{9nt}{2}$ and $v = 3n$. Consequently, the qubit efficiency of the protocol of Ref.[20] is $\eta = \frac{n}{\frac{9}{2}nt+3n} = \frac{2}{9(32+r+d+\delta)+6}$. From Table 1, it can be concluded that this protocol takes advantages over the protocol of Ref.[20] on the aspects of quantum measurement of TP, privacy of summation result towards TP and practical feasibility.

Table 2  Comparison results of this protocol and the previous semiquantum summation protocol

|  | The protocol of Ref.[20] | This protocol |
|---|---|---|
| Characteristic | measure-resend | measure-resend |
| Number of communicants | three | two |
| Quantum resource | single-qubit states | two-qubit entangled states |
| Quantum measurement of TP | single-qubit measurements and three-qubit entangled state measurements | single-qubit measurements and two-qubit entangled state measurements |
| Quantum measurement of communicants | single-qubit measurements | single-qubit measurements |
| Type of TP | semi-honest | semi-honest |
| TP's knowledge about the summation result | Yes | No |
| Usage of quantum entanglement swapping | No | No |
| Usage of unitary operations | No | No |
| Usage of pre-shared key | No | No |
| Summation type | modulo 2 addition | modulo 2 addition |
| Computation way | bit-by-bit | bit-by-bit |
| Qubit efficiency | $\frac{2}{9(32+r+d+\delta)+6}$ | $\frac{1}{6(4+r+d+\delta)+6}$ |
| Quantum channel | ideal noiseless quantum channel | collective-dephasing noise quantum channel |

In conclusion, in this paper, a two-party secure semiquantum summation protocol is constructed, which needn't require two communicants to possess full quantum capabilities. In this protocol, two classical communicants can successfully calculate the summation of their private binary sequences with the assistance from a quantum semi-honest TP, who is permitted to perform all kinds of attacks at her own will except colluding with anyone else. This protocol needn't pre-share a random key between any two parties. This protocol resists the collective-dephasing noise by adopting logical qubits within decoherent free space as traveling particles. It has been validated that this protocol can overcome both the outside attacks from Eve and the participant attacks from TP. Moreover, this protocol can guarantee that TP has no knowledge about the summation results. In the future, we will concentrate on studying how to design the semiquantum summation protocol feasible over other noisy quantum channels, such as the collective-rotation noise channel, the amplitude damping channel, *etc*. As two classical communicants can know each other's private binary sequence, this protocol cannot be directly applied into SQPC [27,34-38]. In the future, we will also study how to apply semiquantum summation into the other semiquantum secure computation, such as SQPC, semiquantum key agreement (SQKA) [39-41], *etc*.

**Acknowledgments**

The authors would like to thank the anonymous reviewers for their valuable comments that help enhancing the quality of this paper. Funding by the Fundamental Research Funds for the Provincial Universities of Zhejiang (Grant No. JRK21002) and Zhejiang Gongshang University, Zhejiang Provincial Key Laboratory of New Network Standards and Technologies (No.2013E10012) is gratefully acknowledged.

**References**

[1] Heinrich, S.: Quantum summation with an application to integration. J Complex, 2002, 18(1):1-50
[2] Heinrich, S., Kwas, M., Wozniakowski, H.: Quantum Boolean summation with repetitions in the worst-average setting. 2003, http://arxiv.org/pdf/quant-ph/0311036
[3] Hillery, M., Ziman, M., Buzek, V., Bielikova, M.: Towards quantum-based privacy and voting. Phys Lett A, 2006, 349(1-4):75-81
[4] Du, J.Z., Chen, X.B., Wen, Q.Y., Zhu, F.C.: Secure multiparty quantum summation. Acta Phys Sin, 2007, 56(11):6214-6219
[5] Chen, X.B., Xu, G., Yang, Y.X., Wen, Q.Y.: An efficient protocol for the secure multi-party quantum summation. Int J Theor Phys, 2010, 49(11):2793-2804
[6] Zhang, C., Sun, Z.W., Huang, Y., Long, D.Y.: High-capacity quantum summation with single photons in both polarization and spatial-mode degrees of freedom. Int J Theor Phys, 2014, 53(3):933-941
[7] Gu, J., Hwang, T., Tsai, C.W.: Improving the security of 'High-capacity quantum summation with single photons in both polarization and spatial-mode degrees of freedom'. Int J Theor Phys, 2019, 58:2213-2217
[8] Zhang, C., Sun, Z.W., Huang, X.: Three-party quantum summation without a trusted third party. Int J Quantum Inf, 2015, 13(2):1550011
[9] Shi, R.H., Mu, Y., Zhong, H., Cui, J., Zhang, S.: Secure multiparty quantum computation for summation and multiplication. Sci Rep, 2016, 6:19655
[10] Shi, R.H., Zhang, S.: Quantum solution to a class of two-party private summation problems. Quantum Inf Process, 2017, 16(9):225
[11] Zhang, C., Situ, H.Z., Huang, Q., Yang, P.: Multi-party quantum summation without a trusted third party based on single particles.

Int J Quantum Inf, 2017, 15(2):1750010
[12] Yang, H.Y., Ye,T.Y.: Secure multi-party quantum summation based on quantum Fourier transform. Quantum Inf Process, 2018, 17(6): 129
[13] Ji, Z.X., Zhang, H.G., Wang, H.Z., Wu, F.S., Jia, J.W., Wu, W.Q.: Quantum protocols for secure multi-party summation. Quantum Inf Process, 2019, 18:168
[14] Duan, M.Y.: Multi-party quantum summation within a d-level quantum system. Int J Theor Phys, 2020, 59(5):1638-1643
[15] Ye, T.Y., Hu J.L.: Quantum secure multiparty summation based on the mutually unbiased bases of d-level quantum systems and its application (in Chinese). Sci Sin-Phys Mech Astron, 2021,51(2): 020301
[16] Ye, T.Y., Hu J.L.: Quantum secure multiparty summation based on the phase shifting operation of d-level quantum system and its application. Int J Theor Phys, 2021, 60(3): 819-827
[17] Bennett, C. H., Brassard, G.: Quantum cryptography: public-key distribution and coin tossing. In: Proceedings of the IEEE International Conference on Computers, Systems and Signal Processing. Bangalore: IEEE Press, 1984, 175-179
[18] Boyer, M., Kenigsberg, D., Mor, T.: Quantum key distribution with classical Bob. Phys Rev Lett, 2007, 99(14):140501
[19] Boyer, M., Gelles, R., Kenigsberg, D., Mor, T.: Semiquantum key distribution. Phys Rev A, 2009, 79(3): 032341
[20] Zhang, C., Huang, Q., Long, Y.X., Sun, Z.W.: Secure three-party semi-quantum summation using single photons. Int J Theor Phys, 2021, 60: 3478-3487
[21] Li, X .H., Deng, F.G., Zhou, H.Y.: Efficient quantum key distribution over a collective noise channel. Phys Rev A, 2008, 78: 022321
[22] Walton, Z.D., Abouraddy, A.F., Sergienko, A.V., et al.: Decoherence-free subspaces in quantum key distribution. Phys Rev Lett, 2003, 91: 087901
[23] Zhang, Z.J.: Robust multiparty quantum secret key sharing over two collective-noise channels. Physica A, 2006, 361:233-238
[24] Gu, B., Mu, L.L., Ding, L.G., Zhang, C.Y., Li, C.Q.: Fault tolerant three-party quantum secret sharing against collective noise. Opt Commun,2010,283:3099-3103
[25] Yang, Y.G., Xia, J., Jia, X., Zhang, H.: Comment on quantum private comparison protocols with a semi-honest third party. Quantum Inf Process, 2013, 12(2):877-885
[26] Zhang, M.H., Li, H.F., Peng, J.Y., Feng, X.Y.: Fault-tolerant semiquantum key distribution over a collective- dephasing noise channel. Int J Theor Phys, 2017, 56: 2659-2670
[27] Lin, P.H., Hwang, T., Tsai, C.W.: Efficient semi-quantum private comparison using single photons. Quantum Inf Process, 2019,18:207
[28] Cai, Q.Y.: Eavesdropping on the two-way quantum communication protocols with invisible photons. Phys Lett A, 2006, 351(1-2): 23-25
[29] Gisin, N., Ribordy, G., Tittel, W., Zbinden, H.: Quantum cryptography. Rev Mod Phys, 2002,74(1):145-195
[30] Deng, F.G., Zhou, P., Li, X.H., Li, C.Y., Zhou, H.Y.: Robustness of two-way quantum communication protocols against Trojan horse attack. 2005, http://arxiv.org/pdf/quant-ph/0508168.pdf
[31] Li, X.H., Deng, F.G., Zhou, H.Y.: Improving the security of secure direct communication based on the secret transmitting order of particles. Phys Rev A, 2006,74:054302
[32] Gao, F., Qin, S.J., Wen, Q.Y., Zhu, F.C.: A simple participant attack on the Bradler-Dusek protocol. Quantum Inf Comput, 2007, 7:329
[33] Cabello, A.: Quantum key distribution in the Holevo limit. Phys Rev Lett, 2000, 85:5635
[34] Chou, W.H., Hwang, T., Gu, J.: Semi-quantum private comparison protocol under an almost-dishonest third party. 2016, https://arxiv.org/abs/1607.07961
[35] Thapliyal, K., Sharma, R.D., Pathak, A.: Orthogonal-state-based and semi-quantum protocols for quantum private comparison in noisy environment. Int J Quantum Inf, 2018, 16(5): 1850047
[36] Ye, T.Y., Ye, C.Q.: Measure-resend semi-quantum private comparison without entanglement. Int J Theor Phys, 2018, 57(12):3819-3834
[37] Lang, Y.F.: Semi-quantum private comparison using single photons. Int J Theor Phys, 2018, 57(10): 3048- 3055
[38] Zhou, N.R., Xu, Q.D., Du, N.S., Gong, L.H.: Semi-quantum private comparison protocol of size relation with d-dimensional Bell states. Quantum Inf Process, 2021, 20:124
[39] Shukla, C., Thapliyal, K., Pathak, A.: Semi-quantum communication: protocols for key agreement, controlled secure direct communication and dialogue. Quantum Inf Process, 2017, 16: 295
[40] Zhou, N.R., Zhu, K.N., Wang, Y.Q.: Three-party semi-quantum key agreement protocol. Int. J. Theor. Phys. 2020, 59(3): 663-676
[41] Li, H.H., Gong, L.H., Zhou, N.R.: New semi-quantum key agreement protocol based on high-dimensional single-particle states. Chin Phys B, 2020, 29(11): 10304